# Development of the Preparation Session Observational Tool (PSOT) for Quantitative Analysis by Probing the Underlying Behaviors that Contribute to Various Learning Assistant-Faculty Relationships

By Fidel Amezcua

# Table of Contents

# Abstract

In order to have a successful Learning Assistant Program it is important to create adequate support tools and resources to adopt and implement the model. The Learning Assistant Model was created at the University of Colorado – Boulder and consists of three essential elements (pedagogy, content, and practice). Our research focuses on the content component, or the weekly content preparation session, in which faculty and their Learning Assistants (LAs) meet to discuss course objectives, content, pedagogical approaches, and student learning. For this study, video recordings of weekly content preparation sessions of various LA-faculty assignments at Chicago State University were analyzed. This data and a review of previous research on observation protocols led us to develop the Preparation Session Observation Tool (PSOT). PSOT is used to develop a better understanding of how the weekly content preparation sessions are fostering different types of partnerships. It is expected that PSOT data can be used by new adoptees of the LA Model and experienced members of LA Programs to analyze their weekly content preparation session in order to improve or change LA-faculty working relationships.

# Introduction

The movement towards reformed education practices in STEM has allowed many undergraduate institutions to change the way traditional science courses are being taught. The Colorado Learning Assistant Model, a product of reformed education, supports innovative research-based instructional practices that benefits students in LA supported classes, LAs themselves, and faculty. It is within this context in which high performing students or Learning Assistants can closely interact with faculty to further their personal understanding of educational and lifelong success. Instructors who partake in or wish to adopt the LA Model should understand that they not only have influence on their students during class time, but they may have a deeper influence on their LAs by ultimately directing them into professional endeavors that would otherwise seem unattainable. In addition, LAs can provide input on instruction that can foster more effective learning environments. It is therefore vital to study how these close interactions between faculty and LAs unfold to better understand how the LA Model can positively impact its participants; especially the non-traditional students who have been historically underrepresented. The tool we developed, to study these interactions, can assist LAs and faculty in better understanding and modifying the type of instructional partnership they are involved in.

The Learning Assistant Program implemented in 2010 at Chicago State University was modeled after the Colorado Learning Assistant Model created at the University of Colorado - Boulder. The LA Model was originally created to deal with shortcomings in U.S. Education. Specifically, the lack of adequate teacher preparation for high school physics teachers throughout the nation.[1] The model also supports active engagement instructional practices and research-based instructional materials in STEM college classrooms which have been found to promote student understanding.[2] Ultimately, the Colorado Learning Assistant Program played a role in redesigning courses in order to accomplish four goals. These four goals include course transformation to enhance the education of science and mathematics students (K-16), increase the recruitment of potential teachers and improve their career preparation, employ discipline-based educational research methods to assist faculty in the preparation of future instructors, and change the culture of science departments to value research-based teaching.[3]

# Background

## *The Learning Assistant Model*

The Colorado Learning Assistant Model is composed of three essential elements which include pedagogy, weekly content preparation, and practice. [3]

### The Pedagogy Course

New LAs are required to take a pedagogy course which introduces them to science education theory and practice.[3] At CU - Boulder, LAs engage in a specific course in the School of Education titled Mathematics and Science Education.[3] The course was specially created to complement and assist LA experiences in their current and future duties at their respective institutions. Because of cultural, population and resource differences the pedagogy course at CSU was developed to build on CSUs strengths and provide specific opportunities to LAs. At CSU the pedagogy course, or the Teacher Immersion Course, was created through the collaborative efforts of CSU faculty and local high school in-service teachers. The model's main focus is to develop competent and reform-oriented instructors (LAs) by focusing on early opportunities for potential teachers to learn and perform reform-oriented instruction, understand discipline-specific instruction, and have access to mentorship from HS instructors who have experience and understand successfully implemented reform-oriented teaching at local high-school institutions. In addition, the Teacher Immersion Course allows undergraduate students to experience lesson planning, lesson assessing, and implementing a single science lesson while being assisted by high school instructors.[4] Practice occurs when LAs finally employ their specific lessons that they develop during class in either the high school or the college setting. LA duties include, but are not limited to, assisting working groups through an inquiry-based approach, helping the instructor hand out worksheets, and periodically checking in on student content understanding.

The Teacher Immersion Course at CSU also allows new LAs to connect their understanding of reform-oriented instructional practice to their LA assignments. The course syllabus has three main objectives and outcomes which allow LAs to understand and reflect on 1) the learning process of STEM students on science related concepts 2) creating research-based instructional materials through the assessment of student knowledge 3) creating a classroom activity that promotes active engagement for either high school or college students. Furthermore, students taking the Teacher Immersion Course assess each objective and outcome by 1) reading literature on education research related to student understanding and writing a two page paper, 2) conducting an informal interview with STEM students at CSU, and 3) evaluating if the classroom activity created effectively promoted student understanding. These objectives and assessments are also facilitated throughout the semester by other in-class activities. For example, students engage in a series of activities that include reading articles about student learning and pedagogical approaches for teaching science related topics, watching videos of LAs applying their teaching and content expertise during in-class interactions with students, and engaging in whole class discussions about the importance/difference between open-ended and closed-ended questions during LA duties.

### The Weekly Content Preparation

The content portion of the LA Model, also known as the weekly content preparatory sessions, allows for faculty and LAs to meet once a week to reflect on student understanding, engage in student assessment data, and prepare for upcoming class work. These weekly content preparatory sessions also differ from one institution to the next. At CU - Boulder, the weekly content preparatory session also includes Teaching Assistants (TAs) alongside the LAs and faculty members.[3] Meetings at CU - Boulder often involve multiple LAs and TAs working with LA Program coordinators. In contrast, at smaller institutions, like CSU, these meetings usually take place in one-on-one settings between the faculty member and the LA. It is the social context in which these weekly content preparatory sessions occur that is of special interest for this research.

Previous research conducted at CSU shows that it is within these spaces that three different types of partnerships evolve between the LA and the faculty member. The partnerships include: mentor-mentee, faculty-driven collaboration, and collaborative. Mentor-mentee is one-directional in which the faculty member focuses more on reviewing content with the LA and going over labs for the upcoming week. Faculty-driven collaboration occurs during scenarios in which the faculty accepts feedback from LAs, yet LAs are not given the freedom to co-design any new instructional materials. Lastly, collaborative partnerships occur when faculty accept feedback from the LAs, and allow the LAs to co-create instructional materials. These partnerships arise due to the close interactions between both parties.[5]

### Practice

Because Chicago State University and the University of Colorado differ in many ways, i.e. student population, class size, demographics, type of institution (R1 University vs. Comprehensive University), and student-to-teacher ratio, it is evident that the LA Program and its various components at CSU will differ from the LA Program at CU - Boulder. At CSU course sizes are significantly smaller than those found at larger institutions. Hence, only one LA is typically assigned to a single course. In contrast, at CU - Boulder there can be multiple LAs per course.[3] Being that these settings are very different, the focus of this study is on the weekly content preparation sessions at our home institution and involves quantitative and qualitative measures to determine characteristics of the weekly content preparation session.

## *Instruments that led to the Development of PSOT*

National calls and advances for reformed, evidenced-based STEM instructional practices at the college level have been a catalyst for institutional change in recent years. Education research projects supported by the National Science Foundation (NSF) since 1991 have led to an increase in studies of student centered instruction.[6] National initiatives such as the Widening Implementation and Demonstration of Evidence-Based Reforms program (NSF), the STEM Education Initiative (Association of American Universities), and the current NSF Improving Undergraduate STEM Education (IUSE) Program have called for increased adoption of evidenced-based instructional practices and training of STEM faculty.[6,7] These programs allow the transformation of traditional lectures, recitations, and laboratory courses into more student-centered courses.[6] The initial impacts of reformed STEM education during the mid-90s, however, were difficult to measure with the tools accessible at that time. New evaluation frameworks and data collection instruments were in need that measured the degree to which these reformed courses coordinated with reformed criterion.[7]

A number of observational instruments that measure the effectiveness of reform-based instructional practices presently exist. These observational tools include, but are not limited to, the Reformed Teaching Observation Protocol (RTOP), the Classroom Observation Protocol for Undergraduate STEM (COPUS), COPUS profiles, and the Laboratory Observation Protocol for Undergraduate STEM (LOPUS).[7–10] Although these tools have their limitations, they have proven to be essential in documenting and providing evidence for reformed-practices in secondary and higher education settings.

We describe the RTOP and COPUS in some detail since these two tools were instrumental in helping us develop our tool, the PSOT, which we utilize in the analysis of the weekly content preparation sessions. Both observational protocols are well established in education research. These tools are well structured, and provide coding schemes to inform the nature of the classroom through the collection of quantitative data. While the RTOP was used to inform expected behaviors as well as our coding, COPUS was used to structure our observational logistics.

Using relevant literature focused on instructional reform to establish a model for reformed teaching, the RTOP was created to quantitatively measure reformed instruction and learning.[7] The observational tool consists of three major scales which are classified as Lesson Design and Implementation, Content, and Classroom Culture. Two of the scales are further divided into subscales. For example, the subscales of the Content component are titled Procedural/Pedagogic Knowledge. The subscales for the Classroom Culture component are titled Communicative Interactions and Student/Teacher Relationships. Each category consists of five likert-type scale items totaling 25 items.[7,9] Interrater reliability, which measures individual raters frequency of differences between responses, is used as one measure of the quality of the instrument.[11] In other words, interrater reliability is a measure of the level of agreement and disagreement between two independent researchers/coders for a section of data. Overall, RTOP requires sufficient training for observers to obtain good interrater reliability as well as make appropriate holistic judgments for the three main categories.[7]

By testing various observation protocols and modifying the format, procedure, data structure, and coding strategies, COPUS was created to ease the process of collecting data on general teaching practices in STEM courses and student/teacher behaviors as well as identify professional development needs for faculty at the undergraduate level.[8] The observational tool consists of 25 behavioral codes in two categories (What the Students are Doing/What the Instructor is Doing) which do not require the observer to make judgments on teaching quality.[8] In addition, COPUS documents classroom behaviors in 2-min intervals through the duration of a class period. Analysis of the observational codes produces two pie charts for each category which show the prevalence of each code. Prevalence of specific behaviors is calculated by dividing the total number of occurrences of each individual code used by the total number of overall used codes.[9] Ultimately, COPUS observers of different professional backgrounds can obtain high interrater reliability and easily implement the observational tool with 2 hours of training.[8,9]

To further gauge reformed practices at the undergraduate level, educational researchers used both RTOP and COPUS to create COPUS profiles. Cluster analysis of 269 (79 faculty form 28 different institutions) recordings for different class periods, i.e. data sets, identified the eight most descriptive, non-redundant, COPUS codes. Cluster analysis groups objects or observations into similar sets or groups through statistical measures.[12] For the researchers of the COPUS profiles, the goal of the cluster analysis was to sort individual class periods into somewhat homogenous clusters given a set of COPUS codes.[9]

Further cluster analysis of the eight codes identified 10 specific instructional strategies which fall in four general instructional styles.[9] To further characterize classroom practices, the 10 COPUS profiles were analyzed together with the RTOP scores. Ultimately, the 10 COPUS profiles are representative of various general instructional styles present in college level STEM courses. A rubric was created for COPUS profiles which summarize specific code percentages within each instructional type. It is important to note that the COPUS codes were analyzed through the prevalence of each code as a percentage of time periods.[9] This allowed for cross comparisons between classrooms which was essential in creating the 10 COPUS profiles.

As part of the national effort to incorporate evidence-based reformed instruction, it has been proposed that using LAs can aid in course transformation since LAs assist in student-centered and peer interactions.[13] Therefore, it is of great interest to investigate the Learning Assistant Model through the lens of reformed instruction and learning. As previously mentioned, the Learning Assistant Model consists of three specific components: pedagogy, content, and practice.

### *Introduction to PSOT*
To complement our previous work, detailing the three types of partnerships (mentor-mentee, faculty-driven, and collaborative) between faculty and LA, and further develop a better understanding of how the weekly content preparation sessions were fostering different types of partnerships, we took ideas from COPUS and RTOP to generate a new instrument called the **Preparation Session Observation Tool (PSOT)**. PSOT characterizes the behaviors of LAs and faculty. PSOT provides a quantitative analysis of qualitative data (videos of weekly content preparation sessions). The nature of this research and the analysis of these specific behaviors were novel and innovative. In these contexts, the existing observational protocols could not assist in the analysis of LA/faculty interactions and potential LA/faculty partnerships. To conduct this work and generate useful findings for practitioners and researchers, a new observational protocol had to be created.

PSOT allowed for a finer-grained analysis of LA/faculty partnerships that goes beyond our three broad partnership categories.[14] The tool was created to observe, characterize, and interpret LA-Faculty behaviors which lead to the different types of partnerships that develop during the weekly meetings. Due to previous work conducted using COPUS and COPUS profiles, as well as knowledge of RTOP, it was decided to use the data gathering structural format of COPUS. This meant analyzing each two minute section of a video episode. The structure was selected because (1) it serves to characterize classroom behaviors of both instructors and students during class time, (2) it was used to create COPUS profiles and (3) it provides a fine grained analysis of specific interactions through an objective lens.

As a tool for analysis, the PSOT is objective. PSOT codes differ from COPUS codes because the types of behaviors observed vary. The environment we are studying is also different: specifically, looking at LA weekly content preparation sessions (PSOT) vs. classroom instruction (COPUS).

The PSOT consists of 22 observational codes in two main categories (Faculty to LA and LA to Faculty). The researcher documents behaviors for every 2-min interval. Data from PSOT can be represented in two different ways after collection. Overall, The Preparatory Session Observational Tool (PSOT) was created to observe, characterize, and interpret LA-Faculty behaviors that occur during weekly prep sessions at institutions with small LA Programs where one-on-one LA/faculty settings occur, such as Chicago State University.

# Methods

### Study Context/Participants

The weekly content preparation session video recordings used for this study were collected to provide a finer grained analysis and inform our previous research findings.[5] The study focused on a specific case of an LA-faculty assignment at Chicago State University. Specifically, the methods employed in the study suggested that partnerships which range from mentor-mentee to collaborative can develop during these weekly content preparation sessions.[5] Interviewees were selected on the basis of availability. The focus of this study was to capture a case of common behaviors across multiple weekly content preparation session meetings.

### Data Collection

Video recordings of weekly content preparation sessions of various LA-faculty assignments were collected between the 2015 spring semester and 2018 spring semester at CSU. The assignments selected were not discipline specific because the focus of the study was to observe common behaviors instead of teaching practices and any other items related to specific course content. While weekly content preparation sessions vary in length, the LA Alliance suggests that these meetings last an hour. The video recordings collected showed that the amount of time allocated to each meeting by each faculty member varied. However, the time allocated for weekly meetings by individual faculty does not negatively affect the study because PSOT measures time and codes as a percent of the total time spent.

For data recording, eight faculty members and nine LAs, representing three different disciplines were selected. Table 1 illustrates the semesters in which individuals were recorded along with their respective disciplines. With the exception of two faculty member who participated for multiple semesters, all other faculty participated for only one semester. 12 video recordings of entire weekly preparation session episodes were obtained. While some of the video recordings were obtained prior to the start of the research, during the 2017 spring semester a faculty likert-scale survey was distributed to all faculty participating in the LA program to obtain background information on each instructors intended use of their LAs. The survey data was used to select and record instructors who utilized LAs in diverse ways to further increase the number of video recordings in our database. IRB protocol guidelines were followed for all interviews, surveys, and video recordings during the data collection process.

**Table 1.** Illustration of the sequence in which faculty and LAs were recorded: each paring is associated with their respective discipline, and the total number of recordings obtained per group.

| Semester | Faculty | LA | Discipline | Total Recordings |
|---|---|---|---|---|
| Spring 2015 | FA1/FA2 | LA1 | Physics | 2 |
| Spring 2016 | FA3 | LA2 | Chemistry | 2 |
| Fall 2016 | FA4/FA5 | LA3/LA4 | Chemistry | 1 |
| Spring 2017 | FA4 | LA5 | Chemistry | 2 |
| Spring 2017 | FA6 | LA6 | Biology | 1 |
| Fall 2017 | FA7 | LA7 | Chemistry | 1 |
| Spring 2018 | FA7 | LA7 | Chemistry | 1 |
| Fall 2018 | FA4 | LA8 | Chemistry | 1 |
| Fall 2018 | FA8 | LA9 | Biology | 1 |

*Video Analysis & Development of PSOT Codes*

The development of observational codes for PSOT occurred through a trial and error process which required multiple iterations, corrections, and repeated analysis of video recordings.

Initially, a rubric scheme of expected general behaviors was created to obtain an idea of LA/faculty behaviors during the weekly content preparation session meetings. The final objective was to compartmentalize these expected behaviors into each type of partnership. These expected behaviors were created by incorporating research data from our previous study.[5] The initial rubric scheme developed for general behaviors specific to each code is shown in Figure 1. Once the initial rubric was completed, video recording observations to define objective common behaviors began.

| Collaborative Partnership | Faculty-driven Collaborative Partnership | Mentor-Mentee Partnership |
|---|---|---|
| Idea sharing<br>• Content delivery<br>• Course organization<br>• Co-development of instructional material | Faculty provides ideas, idea sharing | Faculty provides ideas, lack of idea sharing |
| Faculty and LA both initiate communication | Faculty and LA both initiate communication | Faculty initiates |
| Going over content<br>Consideration of students understanding | Going over content<br>Unknown consideration | Going over content<br>Consideration of LA's understanding |
| Guidance from one another | Guidance from faculty is not received or wanted | Guidance from faculty |
| Relationship that extends past faculty-student (professional level) | Relationship desired by faculty member only | Surface relationship |

**Figure 1.** Rubric scheme created for expected general behaviors specific during the weekly content preparation session meeting.

Analysis of video recordings began during the 2016 fall semester. These observations were conducted by two undergraduate research students with feedback from two faculty members at CSU. As the rubric was implemented, the need for the creation of new codes for observed behaviors was highlighted. These new codes were essential for characterizing the interactions and representing all the behaviors in the interaction. For instance, discussions reflecting on student engagement and off-topic/on-topic conversations were not in the original rubric but were added based on our observations and the need to develop a complete set of codes. Due to limited video recordings in the initial slate of data, literature for observational protocols was explored to determine other possible behaviors.

### How the Literature informed PSOT Development

Some categories on PSOT were informed by the RTOP instrument. The RTOP category, *The Classroom Culture: Communicative Interactions and Student/Teacher Relationships,* was of special interest.[7] This section of the RTOP was chosen since it focused on specific student/teacher interactions during class time. The items selected for reference from this category were items 16, 17, 19, 21, 22, and 24, described below. These items were not used as specific PSOT codes, but were instead used to inform potential behaviors useful for PSOT codes. Items 16 through 19 fall within the subcategory *Communicative Interactions under the Classroom Culture* subcategory of RTOP. Item 16 focuses on the student communication involvement of ideas through various mediums (i.e. media, interpersonal, etc.). Item 17 is intended to observe how teachers allowed students to exhibit different modes of thinking through questions. Item 19 narrows in on how and if the focus and direction of classroom discourse was determined by student questions and comments.[7] Items 21, 22, and 24 fall within the subcategory *Student/Teacher Relationships*. Item 21 documents whether active participation was encouraged for students. Item 22 documents if students were also encouraged to think critically about conjectures, solution strategy, and evidence interpretation. Lastly, item 24 focuses on whether the instructor served as a resourceful person who aided and enhanced student investigations.[7]

The literature on the observational tool, COPUS, proved to be more difficult as a resource for potential observational codes. Observational codes in COPUS are centered on in-class interactions of student/teacher behaviors. Therefore, only two observational codes found in COPUS were directly incorporated into PSOT (Listening and Other). Of special interest, however, was the structure of the tool: the methodology used to create specific COPUS observational codes and the methods used to determine interrater reliability among observers. COPUS captures observational codes at two-minute intervals and also provides graphical data useful for analysis. The same two-minute intervals for graphical data and analysis are used with PSOT.

Observing video recordings in a second round of video analysis generated new ideas for observational codes. At this point new video recordings had been acquired, so video analysis consisted of both new and older recordings. As we watched the video recordings of the weekly content preparation sessions, keeping our RTOP and COPUS reference work, a number of observational codes specific to the PSOT were generated.

Observational codes were created and divided into two categories (Instructor to LA and LA to Instructor). Observational codes were strategically renamed, disregarded, or modified based on repeated viewing and iterative analysis. This resulted in seven iterations of the PSOT. The final two categories for PSOT were also renamed (Faculty to LA and LA to Faculty). 13 codes were created in the category Faculty to LA and 11 codes in the LA to Faculty category totaling 24 codes. Figure 2 shows the final version of the PSOT and figure 3 shows the final version of the 24 PSOT observational codes.

|      | Faculty to LA |    |    |     |    |    |     |     |     |      |       |    |   | LA to Faculty |    |     |     |    |    |    |      |       |    |   |
|------|---|----|----|-----|----|----|-----|-----|-----|------|-------|----|---|---|----|-----|-----|----|----|----|------|-------|----|---|
| Time | L | NT | LT | TGI | TL | FR | FRQ | IMP | IMN | F-IM | F-IMQ | NK | O | L | NT | CTU | CTS | WM | LT | LR | L-IM | L-IMQ | NK | O |
| 0    |   |    |    |     |    |    |     |     |     |      |       |    |   |   |    |     |     |    |    |    |      |       |    |   |
| 2    |   |    |    |     |    |    |     |     |     |      |       |    |   |   |    |     |     |    |    |    |      |       |    |   |
| 4    |   |    |    |     |    |    |     |     |     |      |       |    |   |   |    |     |     |    |    |    |      |       |    |   |

**Figure 2.** Final version of the PSOT: the observational codes acronyms for each category are coded for every two minutes (Time).

**Faculty to LA Observational Codes**

- **L**    Listening: Listening to LA
- **NT**    Neutral Talk: Small talk not connected to course or student understanding for a period longer than 10 seconds
- **TGI**    Teaching Guided Inquiry: Faculty member guides LA through course content by model inquiry teaching
- **TL**    Teaching Lecturing: Faculty member guides (does not talk about content only) LA through course content by lecture
- **LT**    Logistic Talk: Dialogue concerning future course logistics such as class meeting times, office hours, exam dates - class related items
- **FR**    Faculty Reflection: Faculty member describes to LA the students' (in-class) understanding/misconceptions/resources/etc.
- **FRQ**    Faculty Reflection Questioning: Faculty member inquires about in class LA-Student discourse in which an immediate response is obtained from the LA
- **IMP**    In-class Modification Positive: Faculty reacts positively to L-IM (LA In-class Modification)
- **IMN**    In-class Modification Negative: Faculty member reacts negatively to L-IM (LA In-class Modification)
- **F-IM**    Faculty In-class Modification: LA receives instruction on how to create activity/Faculty member develops instructional materials/curriculum or asks LA to do so
- **F-IMQ**    Faculty In-class Modification Questioning: Faculty member elicits feedback/seeks insight from the LA about potential in-class modifications through questioning
- **NK**    Note Taking
- **O**    Other: Any other behavior of interest

**LA to Faculty Observational Codes**

- **L**    Listening: Listening to Faculty
- **NT**    Neutral Talk: Small talk not connected to course or student understanding for a period longer than 10 seconds
- **CTU**    Content Talk Understanding: LA discusses course content and demonstrates understanding
- **CTS**    Content Talk Seeking: LA discusses course content, does not demonstrate understanding, and seeks clarification from faculty member
- **WM**    Working Model: Working on a lab model (physical, conceptual, etc.)
- **LT**    Logistic Talk: Dialogue concerning future course logistics such as class meeting times, office hours, exam dates - class related items
- **LR**    LA Reflection: LA reflects on students' (in-class) understanding/learning and describes their perspective to the faculty member
- **L-IM**    LA In-class Modification: LA suggests changes to class activities/materials/curriculum which can be objectively implied or subtlety/indirectly suggested (occurs by default)
- **L-IMQ**    LA In-class Modification Questioning: LA elicits feedback/seeks insight from the faculty member about potential in class modifications suggested by the instructor through questioning
- **NK**    Note Taking
- **O**    Other: Any other behavior of interest

**Figure 3.** Descriptions of the final LA and Faculty RTOP observation codes

### Examples of PSOT Codes

An example of two codes specific to PSOT will be used to help the reader understand how the codes relate to the behaviors observed. The codes of interest are the LA & Faculty suggested in-class modifications. This example below is a portion of a transcript from a dialogue between an LA and a faculty member during a weekly content preparation session supporting a biochemistry course and illustrates a collaborative partnership.[14] The following example shows a suggested in-class modification by the LA in which the LA attempts to reinforce understanding and make the conceptual load more manageable for an upcoming in-class activity. The faculty member then follows up and decides to incorporate the suggestion from the LA.[14] The following is a section of transcript that would be categorized as (L-IM/IMP). Note that "LA" represents the LA, "F "represents the faculty member (instructor), "…" indicates that words, that did not change the overall meaning, were omitted from the transcript, and text in "( )" are notes for the reader but were not said during the discussion.

| | |
|---|---|
| LA: I'm thinking even with glycolysis or the glucose in the entire processes…it makes we wonder if we can go through the top half of that, but I think once again that probably depends on the time that we are working with, so it won't seem…overwhelming…you know, all of these enzymes or all of this information…maybe we can du a, uh, small like mini-quiz or recap prior… | L-IM |
| F: Yes. OK - OK! (writes down on her notes) | IMP |
| LA: …that way it won't seem extremely heavy like 'oh my gosh we have to remember all of these different-you know, enzymes and - | L-IM |
| F: So a little, umm, mini-quiz? | F-IMQ |
| LA: Yeah, I think that would be good…just to kinda see…what they should have harnessed… | |
| F: You think they'll like that? | F-IMQ |
| LA: I think so, because it gives them the opportunity to not only go through the information and process it but actually go home and study…to see how that's going to play into-you know, learning the second half. | |

In this example, the LA presents the possibility of quizzing the students immediately after they have covered material to assess the student's memory and understanding (L-IM). After some consideration the instructor decides to incorporate the activity into the course (IMP).[14] These two codes are suggestive of a collaborative partnership because this faculty member allowed for a comfortable environment in which LAs can provide suggestions that can potentially change the structure of a lesson.

***Validity and Reliability of the Tool***

The IRR statistical analysis used for PSOT is the same as the one used to validate COPUS.[8] Because both the RTOP and COPUS code for categorical data, Cohen's Kappa is used to measure IRR since it accounts for the possibility that observers agree by chance.[9] For our analysis Microsoft Excel was used to calculate Cohen's Kappa values instead of SPSS. In terms of validity, the behaviors documented by PSOT are representative of the weekly content preparation portion of the LA Program at CSU. As a model, it is well established that all three aspects of the LA model overlap to support one another and can influence the types of partnerships that evolve. Therefore, it is important to note that PSOT provides an analysis of the weekly content preparation session only. There is no guarantee that discussions and plans made between the LA and faculty member during the weekly content preparation session make it into the classroom practice. Analysis of the practice portion of the LA Model would be worthwhile follow up but is beyond the scope of this work.

It is essential to achieve a high level of IRR when PSOT is used. To obtain a valid level of IRR, student-researchers at CSU independently coded 10 video recordings. Due to the vagueness of some codes however a high level of IRR was not obtained during initial work. Thereafter, both student-researchers compared and identified differences in coding which led to a second round (three videos) of coding in which the student observers sat next to each other and checked-in at 5-minute intervals to compare codes. Average Cohen's Kappa values for the first round of coding was 0.636. After the individual coders came to a consensus about the specificity of the codes, an average Cohen's Kappa value of 0.859 was obtained for the second round of coding. Table 1 shows the ranges of Cohen's Kappa values and their associated strength of agreement used to rate our observed IRR values.[9]

**Table 2.** Ranges of Cohen's Kappa values and their associated level of agreement used to rate our observed IRR values.[15]

| Value of Kappa | Level of Agreement | % of Data that are Reliable |
|---|---|---|
| 0 – .20 | None | 0 – 4% |
| .21 – .39 | Minimal | 4 – 15% |
| .40 – .59 | Weak | 15 – 35% |
| .60 – .79 | Moderate | 35 – 63% |
| .80 – .90 | Strong | 64 – 81% |
| Above .90 | Almost Perfect | 82 – 100% |

## Analyzing PSOT Data

Since PSOT has a similar structure to COPUS, data is analyzed in a comparable fashion. For example, PSOT data can be represented graphically for each weekly content preparation session. The prevalence of different codes is determined by adding up how often each code was observed and then dividing by the total number of codes.[8] This data can then be represented by a pie graph showing percentages of the total codes. The prevalence of PSOT codes can also be analyzed as a percentage of time. Each individual behavioral code checked throughout the observation is divided by the total number of time intervals observed. This data can be represented by a percent bar graph with respect to time intervals.[9]

Once observations have been conducted using PSOT, data can be used to suggest the types of ongoing LA/faculty partnerships because PSOT behavioral codes have been generated to correlate with a specific partnership. The relationship between PSOT behavioral codes and the three faculty partnerships can be observed in figure 3. This representation was developed by Felicia Davenport with modifications by Fidel Amezcua (author). As a visual aid, a graphical representation of a bar graph of percentage with respect to time intervals of a mentor-mentee partnership and a faculty-driven collaboration are shown in figure 4 and figure 5. This graphical data, however, should not be interpreted as a benchmark for identifying these partnerships but as a reference point to compare. Ultimately, identifying the type of partnerships between faculty and LAs will require observers to use their personal judgments when interpreting PSOT data. For instance, if a PSOT analysis is done and there are many instances of L-IM occurring, this may be indicative of a collaborative partnership for some, but others might desire more evidence of collaborative behaviors. We do not operationally define what a collaborative relationship is.

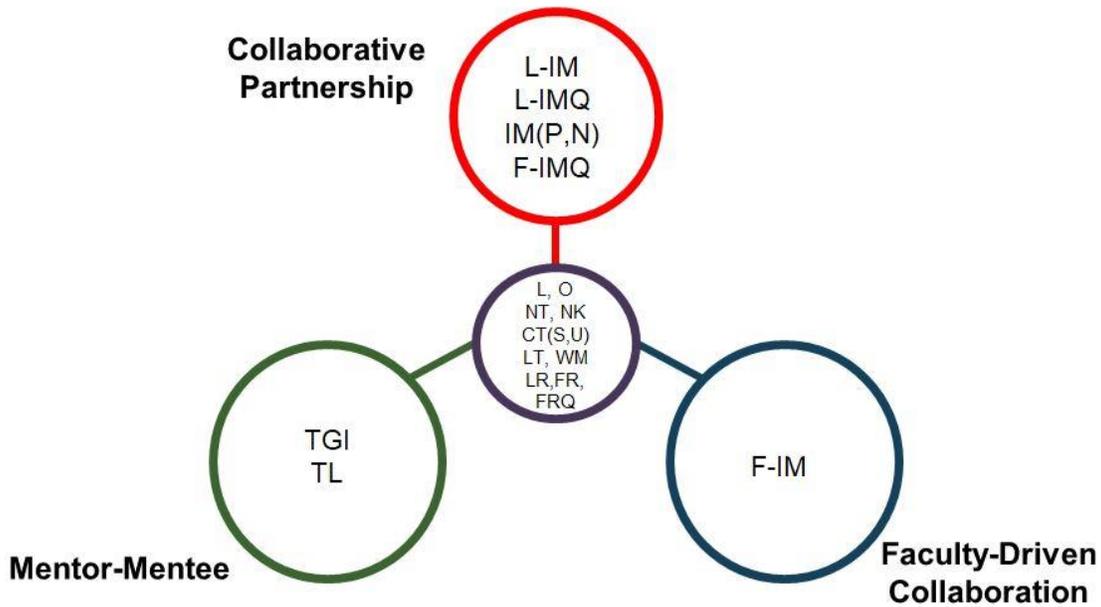

**Figure 3.** shows the relationships between the different LA/faculty partnerships and the PSOT behavioral codes[16]

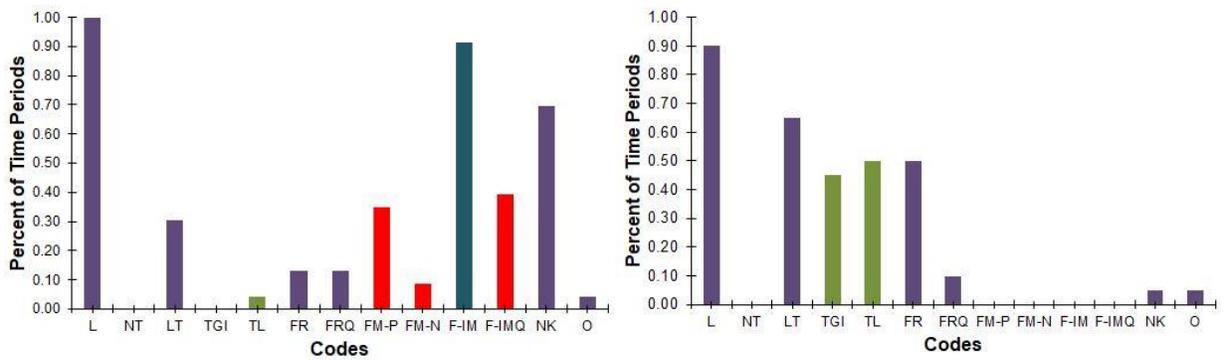

**Figure 4.** This figure displays Faculty to LA observational bar graphs of two different weekly content preparation sessions held by two different CSU faculties. The bar graph on the left illustrates results for a collaborative partnership while the right illustrates results for a mentor-mentee collaboration. The bars have been color coded to match the code color patterns displayed in figure 3.[16]

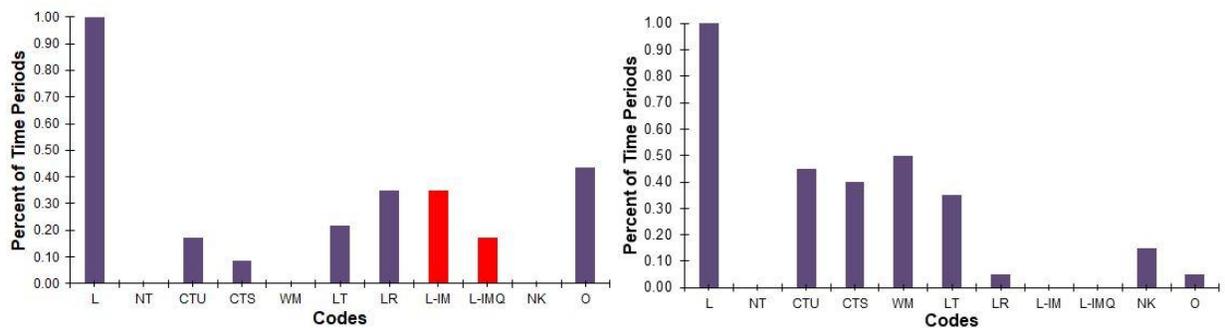

**Figure 5.** This figure displays LA to Faculty observational bar graphs from the same data used for the previous figure. The bar graph on the left illustrates results for a collaborative partnership while the right illustrates results for a mentor-mentee collaboration. The bars have been color coded to match the code color patterns displayed in figure 3.[16]

# Discussion

     While all three components of the LA Model play a significant role in reforming education, transforming classroom culture, and are worthwhile investigating, this study focuses on the weekly content preparation portion of the LA Model. Across institutions the weekly preparation session tends to be the least structured element of the LA Model. For example, large institutions like CU - Boulder incorporate weekly meetings that include multiple LAs, TAs, and faculty while smaller institutions like CSU usually hold a one-on-one weekly meeting between LAs and faculty. Although these weekly meetings are often structured to focus on content, they also provide a platform for cogenerative dialogue, which is the process in which a small number students alongside co-teachers assess in-class evidence and together generate alternative resolutions regarding new class rules, student/teacher roles, and responsibility for establishing these changes, to occur which allows for faculty and LAs to contribute in meaningful ways thus establishing various types of working partnerships.[5,17]

     The need for a finer grained analysis of these working partnerships required a new observational tool. Two observational tools used to identify reformed education practices were used to create the PSOT. PSOT was created to assess the three types of LA-faculty partnerships previously explained and characterized at CSU.[5] Specifically, the RTOP and COPUS were used as starting reference points which led to the creation of 24 observational codes. The codes provide useful information that can be used "by LAs, coordinators, instructors, and researchers with differing expertise in analysis to reflect on their specific instructional partnerships and how they conduct preparation sessions."[14] Additional statistical tools should also be used to verify IRR by observers to check for proper use of the tool. Cohen's Kappa IRR values allow us to validate observations, so if IRR is not good enough (below 0.61 for our purposes) further training should be sought out. Through an iterative process of tool development that I describe in detail in this thesis, we were able to get an average IRR value of 0.859 using the final version of PSOT which is at a strong level of agreement.

     Due to the nature of this study, PSOT is recommended to be used by faculty at smaller institutions which tend to have one-on-one weekly content preparation session meetings. It would be worthwhile assessing whether the tool can be used in weekly content preparation sessions in larger LA programs. However, this is beyond the scope of this study. For larger institutions we recommend researchers assume that observational codes are representative of the variety of behaviors from all LAs, TAs, and faculty present. However, if TAs are responsible and take on the role of the faculty during the weekly content preparation session then codes should be representative of the TAs and not the faculty member. If TAs and instructors hold their own weekly content preparation session, than PSOT can be used to gauge whether 1) it can be

used for that specific scenario 2) assess whether TAs take the instructors role and apply their recommendations during the weekly content preparation sessions or 3) provide opportunities for LAs to establish/develop collaborative partnerships during these weekly content preparation sessions that can potentially create changes in classroom culture. If these weekly content preparation sessions are held in the presence of LAs, TAs, and faculty, then the nature of the weekly content preparation session should be surveyed: i.e. do TAs behave more like faculty or LAs. In its current form, the PSOT assumes that in-class modifications suggested by LAs, and accepted by faculty, are then employed during class time. Additional research should gauge whether these modifications are actually applied to the classroom setting by both the LA and faculty. The data can provide further information on the types of partnerships between the LA and faculty, and how they are implemented during class time.

## Conclusion and Implications

Currently, it is expected that PSOT data can be used at small institutions by new adoptees of the LA Model and experienced members of LA Programs to analyze their weekly content preparation session in order to improve or change their working relationships to the type of partnership desired. By looking at trends in PSOT data, faculty can look to engage in specific activities during their weekly content preparation sessions to create opportunities for LAs to display their content knowledge, incorporate ideas to the discussion, and partake in leadership roles during class time that may facilitate the transformation of the course and class structure. On a departmental level, PSOT can lead to targeted professional development at an institution's LA Program to promote the maximum benefit of faculty, LAs, and students in LA-supported classes. Ultimately, PSOT can be used to effectively establish inclusive learning environments which utilize LAs in diverse ways to provide instructor and student support.[14]